\documentclass[aps, prd, twocolumn, superscriptaddress, nofootinbib]{revtex4}
\usepackage{graphicx}
\usepackage{dcolumn}
\usepackage{amssymb}
\usepackage{amsmath,amssymb,amsfonts}

\begin{document}

\newcommand{\Eq}[1]{\mbox{Eq. (\ref{eqn:#1})}}
\newcommand{\Fig}[1]{\mbox{Fig. \ref{fig:#1}}}
\newcommand{\Sec}[1]{\mbox{Sec. \ref{sec:#1}}}

\newcommand{\PHI}{\phi}
\newcommand{\PhiN}{\Phi^{\mathrm{N}}}
\newcommand{\vect}[1]{\mathbf{#1}}
\newcommand{\Del}{\nabla}
\newcommand{\unit}[1]{\;\mathrm{#1}}
\newcommand{\x}{\vect{x}}
\newcommand{\y}{\vect{y}}
\newcommand{\p}{\vect{p}}
\newcommand{\ScS}{\scriptstyle}
\newcommand{\ScScS}{\scriptscriptstyle}
\newcommand{\xplus}[1]{\vect{x}\!\ScScS{+}\!\ScS\vect{#1}}
\newcommand{\xminus}[1]{\vect{x}\!\ScScS{-}\!\ScS\vect{#1}}
\newcommand{\diff}{\mathrm{d}}

\newcommand{\be}{\begin{equation}}
\newcommand{\ee}{\end{equation}}
\newcommand{\bea}{\begin{eqnarray}}
\newcommand{\eea}{\end{eqnarray}}
\newcommand{\vu}{{\mathbf u}}
\newcommand{\ve}{{\mathbf e}}


\title{Vacuum fluctuations in theories with deformed dispersion relations\\
}

\newcommand{\addressImperial}{Theoretical Physics, Blackett Laboratory, Imperial College, London, SW7 2BZ, United Kingdom}
\newcommand{\addressRoma}{Dipartimento di Fisica, Universit\`a ``La Sapienza''
and Sez. Roma1 INFN, P.le A. Moro 2, 00185 Roma, Italia}

\author{Michele Arzano}
\affiliation{\addressRoma}
\author{Giulia Gubitosi}
\affiliation{\addressImperial}
\author{Jo\~{a}o Magueijo}
\affiliation{\addressImperial}
\author{Giovanni Amelino-Camelia}
\affiliation{\addressRoma}

\date{\today}

\begin{abstract}
We examine vacuum fluctuations in theories with modified dispersion relations which represent dimensional reduction at high energies. By changing units of energy and momentum we can obtain a description rendering the dispersion relations undeformed and transferring all the non-trivial effects to the integration measure in momentum space. Using this description we propose a general quantization procedure, which should be applicable whether or not the theory explicitly introduces a preferred frame. Based on this scheme we evaluate
the power spectrum of quantum vacuum fluctuations. We find that in {\it all}  theories which run to 2 dimensions in the ultraviolet the vacuum fluctuations, in the ultraviolet regime, are scale-invariant. This is true in flat space but also for ``inside the horizon'' modes in an expanding universe. We spell out the conditions upon the gravity theory for this scale-invariance to be preserved as the modes are frozen-in outside the horizon. We also digress on the meaning of dimensionality (in momentum and position space) and suggest that the spectral index could itself provide an operational definition of dimensionality. 
\end{abstract}


\maketitle


\section{Introduction}

The phenomenon of dimensional reduction at the Planck scale has attracted considerable interest in quantum gravity research over the past few years. It is by now well established  that several quantum gravity frameworks, in certain regimes, predict a running of dimensionality towards decreasing values, as smaller and smaller length scales are probed \cite{Ambjorn:2005db,Lauscher:2005qz,Horava:2009if,Modesto:2008jz,Carlip:2009kf,Benedetti:2008gu,Alesci:2011cg,Benedetti:2009ge,Calcagni:2010pa,Magliaro:2009if,Alkofer:2014raa,Calcagni:2014cza,Arzano:2014sya,Arzano:2014jfa,V:2015yma}.  As first noted in the context of causal dynamical triangulations \cite{Ambjorn:2005db}, the notion of ``spectral dimension", associated with a fictitious diffusion process, can provide a useful probe of the ultraviolet (UV) regime of quantum gravity, where the notion of space-time as a smooth manifold may lose its meaning. Interestingly, a running of the spectral dimension to 2 in the UV seems to be a common denominator for various approaches to quantum gravity. The use of the spectral dimension as a genuine indicator of {\it physical} properties of space-time at small length scales has, however, potential drawbacks. For example the heat kernel used to calculate the spectral dimension is only defined for Euclidean spaces, and its trace can assume negative values in some models, leading to obvious problems with its interpretation as a return probability \cite{Calcagni:2013vsa}.

Remarkably, as observed in \cite{Sotiriou:2011aa}, a running spectral dimension can be modelled by a Laplacian whose counterpart in momentum space is a modified dispersion relation (MDR) with deformations triggered by a UV scale. In \cite{DSRmeasure,Amelino-Camelia:2013cfa} we pointed out that by changing energy-momentum units in a way which ``trivializes" the dispersion relation, all the non-trivial features of the model can be encoded in the integration measure in energy-momentum space. Under such a description, the running of the dimensionality translates into the more intuitive running of the {\it Hausdorff dimension} of energy-momentum space, measuring the scaling of the volume with a change of radius. In all cases studied so far such Hausdorff dimension coincides asymptotically with the one obtained using the spectral dimension.

A further notable insight into the physical meaning of running dimensions was pointed out in \cite{DSRflucts2},  where for one concrete theory it was shown how the specific dispersion relations associated with running of dimensionality to 2 lead to a scale invariant spectrum of primordial cosmological fluctuations, via a modified speed of sound~\cite{Magueijo:2008pm}, with no need for inflation. If we map the non-trivial effects of the theory fully onto the integration measure in momentum space, we learn that the scale-invariance of the fluctuations emerges directly as an effect of UV Hausdorff dimensional reduction~\cite{DSRmeasure}. However, the arguments used in \cite{DSRmeasure} to link a scale invariant spectrum to a deformed measure only touched briefly upon the quantization procedure of fluctuations under a deformed energy-momentum integration measure.

In this paper we consider quantum vacuum fluctuations within a vast class of theories with MDRs, with or without introducing a preferred reference frame. We evaluate their power spectrum by adopting units (i.e. choosing a frame) which, as in
 \cite{DSRmeasure}, render the dispersion relations quadratic as usual, shifting all the non-trivial effects to the measure of integration in momentum space. The quantization procedure is then that of a theory with trivial dispersion relations (and so second-order equations, and quadratic action) but a non-trivial measure of integration in energy-momentum space. Thus, we connect directly the evaluation of vacuum fluctuations with the Hausdorff dimension description of dimensional reduction.

In Section~\ref{formalism} we propose a straightforward generalization  of the standard quantization framework, by considering the Dirac delta function associated with the non-trivial measure. Specifically we postulate commutation relations for creation and annihilation operators identical with the usual ones but with a delta function associated with the new measure. In Section~\ref{inner} we show that the procedure can be self-consistently closed with generalizations of the inner product of fields, dual measure in position space, and delta function in position space. The usual identities in canonical quantization are then formally valid once the appropriate replacements are made.

Assuming this quantization procedure, in Section~\ref{vac-flucts} we evaluate the power spectrum of vacuum quantum fluctuations.
By defining the ``covariant'' power spectrum $\overline P(p)$ (i.e. a power spectrum
concomitant with the modified integration measure), we can cast our results in the remarkable general form 
$
\overline P(p)=1.
$
The standard power spectrum $P(p)$, however, varies from  theory to theory and depends on the original MDRs
of the specific theory.  We identify the dimensionless power spectrum ${\cal P}(p)$ and the spectral index $n_S$.
We prove the general result that for all theories under consideration the vacuum fluctuations are scale-invariant
if the momentum-space measure's  Hausdorff dimension runs to 2 in the UV.

In Section~\ref{applications1}  and~\ref{applications2} we then apply these general results to a number of topical theories
that can be fitted into our scheme.
We first  consider Lorentz invariance violating (LIV) theories, which explicitly introduce a preferred frame.
In Section~\ref{HL}  we review the Horava-Lifshitz scenario (studied previously in~\cite{DSRflucts1,muko,DSRflucts2,DSRrainbow,DSRmeasure}), showing how some of its well
known properties and facts, specifically as reworked in~\cite{DSRmeasure}, can be understood within the wider picture
proposed here. In Section~\ref{gtneq0} we examine a generalization of~\cite{Sotiriou:2011aa} where the high energy MDRs
acquire non-quadratic powers of the energy. If such MDRs are interpreted as a
higher order derivative theory this is problematic, but if the effect is shifted to the integration measure
in momentum space and only then a reinterpretation in terms of field theory is made, our formalism can be applied.
We set up the quantization of such a theory and evaluate its vacuum
fluctuations.

In Section~\ref{applications2} we consider theories in which Lorentz symmetry is deformed rather than explicitly broken, in the sense of the ``DSR" proposal \cite{AmelinoCamelia:2000mn, AmelinoCamelia:2000ge, KowalskiGlikman:2001gp, Magueijo:2001cr, Magueijo:2002am, KowalskiGlikman:2002ft}. If all steps are taken consistently with the symmetry deformation, then a preferred frame is not introduced. However this consistency with the symmetry deformation is highly non-trivial and care must be taken with several subtleties regarding the quantization process, which we spell out. Our quantization framework applies equally well to several classes of theories, including what might be called  ``Non-local Lorentz invariant theories'' and ``de Sitter DSR''.  We find that here, too, in all cases for which the MDRs represent dimensional reduction to 2 in the UV the fluctuations are scale-invariant. 
Finally, in Section VI we elaborate on the notion of non-trivial dimensionality in momentum space and in the dual space-time picture. In a concluding Section we discuss the overall meaning of our results.

\section{Quantization under a non-trivial measure}\label{formalism}

Let us consider a theory with measure $d\mu (E,\vect{p})$ in energy-momentum space and let us assume that  the theory is such that the particle dispersion relation is the standard quadratic one. As discussed in later sections of this paper, we could have started from a theory with modified dispersion relation and chosen units such that the dispersion relation reduces to the standard form, with all non-trivial effects shifted to the measure $d\mu (E,\vect{p})$. Such a procedure leads to an interesting reinterpretation of the phenomenon of running of the spectral dimension, since the Hausdorff dimension of the measure in the new units fully captures the asymptotical UV dimension of the theory~\cite{DSRmeasure}.   Here we shall assume that a field theory reinterpretation of the theory is only carried out in the news units, and that quantization is performed only at this stage.

Canonical quantization of a field theory starts from the specification of the quantum states. These are constructed from positive energy solutions of the field equation which, under Fourier transform, are mapped into functions on the positive mass-shell. Thus, after taking the Fourier expansion of a field, $\phi$, the first step in quantization consists in going ``on-shell'', i.e. implementing the replacement:
\be
d\mu(E,\p)\rightarrow d\overline\mu(\p)
\ee
such that for any function $f(E,\p)$:
\be
\int d\mu(E,\p)\delta (\Omega)\theta(E)f(E,\p) =\int d\overline\mu(\p) f(E_{\p},\p)
\ee
where $\Omega(E,\p)=0$ represents the dispersion relation and $E_\p$ is the positive solution to $\Omega=0$.
Then,
\be\label{onshellmu}
d\overline\mu(\p)=\frac{d\mu(E_\p,\p)}{2E_\p}
\ee
where $E_\p=\sqrt{\p^2 +m^2}$ in this case. This solution is also
assumed throughout the integrand.

If the measure and integrand are isotropic we can integrate away the angular coordinates, keeping only
the radial spatial-momentum coordinate. The measure can then be written as:
\be\label{isomu}
d\overline\mu(\p)=4\pi \overline\mu(p)\, dp,
\ee
for 3 spatial dimensions (with a different numerical factor coming from the angular integration if  $D\neq3$).
For the standard theory (measure and dispersion relations) with $D=3$, we have:
\be\label{convisomu}
\overline\mu(p)=\frac{p^2}{(2\pi)^3 2E_p}
\ee
as is well known. This is the so-called covariant on-shell measure, in a  convention adopted by many quantum field theory (QFT) textbooks  (e.g. \cite{ryder}).

\subsection{The fundamental postulate}\label{postulate}
Given that the theory can be represented (to second order) by a Klein-Gordon action, we do not have to contend with higher than second-order derivatives in time, even though such higher order derivatives would be present were we to transition to field theory from  the original representation (i.e. before choosing linearizing units). Therefore, the second quantization procedure should closely mimic the standard one, but accounting for the non-trivial measure.

Let us expand the field operator as:
\be\label{expansion}
\phi(x)=\int d\overline\mu(\p) [a(\p)e^{-ip\cdot x}+a^\dagger (\p)e^{ip\cdot x}].
\ee
This expression is just the standard one in the undeformed case, in a notation used by some authors (e.g.~\cite{ryder}).
Specifically, in the undeformed case, the ``covariant'' measure is given by
\be
d\overline\mu(\p)=\frac{d^3p}{(2\pi)^32E_\p}\,.
\ee
Since we are only interested in evaluating the two-point functions of the theory, we consider the action of $a (\p)$ and $a^\dagger(\p)$ only on the vacuum state. As usual the vacuum $|0\rangle$ is defined from $a(\p)|0\rangle=0$. One particle states are given by  $|\p\rangle = a^\dagger (\p)|0\rangle $.
Our fundamental postulate is that these one-particle states satisfy the following orthogonality relation
\be\label{norm}
{\langle \p'|\p\rangle}=\delta_{\overline \mu}(\p-\p')
\ee
to accommodate the modified measure in momentum space, with the Dirac delta $\delta_{\overline\mu}$ such that:
\be
\int d\overline\mu\, \delta_{\overline \mu}(\p)=1.
\ee
Notice that this agrees with the usual expression of the commutator for creation and annihilation operator when evaluated on the vacuum
\be\label{post}
[a(\p),a^\dagger (\p')]=\delta_{\overline\mu}(\p-\p')\,.
\ee
In the undeformed case
\be
\delta_{\overline \mu} (\p) = (2\pi)^3 2E_\p\, \delta^{(3)}(\p)
\ee
where $\delta^{(3)}(\p)$ is the standard delta function in three dimensions, so that the commutator (\ref{post}) for the creation and annihilation operators reads in ordinary QFT:
\be\label{post0}
[a(\p),a^\dagger (\p')]=(2\pi)^3 2E_\p\, \delta^{(3)} (\p-\p')\; .
\ee
As we shall see, using this convention the commutation relations remain simple when used in theories with MDRs, both for LIV and DSR models. In the LIV case we could have simply postulated (\ref{post}) (acting on a general state); however for DSR it is essential that (\ref{post}) is only applied on the vacuum to circumvent well-known problems  in the multiparticle sector of the theory (see, e.g.~\cite{arzano-marciano,arzanoQFT}). A somewhat different approach, with obvious connections to ours, can be found 
in~\cite{sabine1,sabine2}.

\subsection{Position space commutators and inner product}\label{inner}
Since we have started from a modified measure in {\it momentum} space (and given that our ultimate goal is to compute the
power spectrum of vacuum fluctuations), it made more sense to define the QFT from (\ref{post}), i.e.
in terms of the momentum space
commutators. We can, however, investigate the implications for the more conventional starting point of QFT:
the position space equal time commutation relations. We can also consider the status of  the inner product in field space.
In this subsection we show that with simple definitions for the dual measure and delta function in position space
the procedure can be self-consistently closed. However, we stress that this construction is not needed for the main result of this
paper: the power spectrum of vacuum fluctuations. 

Specifically, using (\ref{post}) and (\ref{expansion}) it is easy to show that:
\be
[\phi(\x,0),\dot \phi(\y,0)]=i\int d\overline\mu(\p)\, 2E_\p e^{i\p\cdot (\x-\y)}.
\ee
(We note that we keep the original coordinates in ${\bf x}$ and ${\bf p}$ in our transform, so that the phase here is always dimensionless.)
This suggests defining a dual delta function in position space from:
\be\label{deltax}
\delta_{\overline\mu_x}(\x)=\int d\overline\mu(\p)\, 2E_\p e^{i\p\cdot \x},
\ee
so that:
\be
[\phi(\x,0),\dot \phi(\y,0)]=i \delta_{\overline\mu_x}(\x - \y).
\ee
With this modified delta function we comply with the usual starting point of canonical quantization.

Given such a delta function, the position space dual measure $d\overline\mu_x(\x)$  can be inferred from the requirement:
\be
\int d\overline\mu_x(\x) \, \delta_{\overline\mu_x}(\x)=1.
\ee
For the standard theory it is easy to show that $\delta_{\overline\mu}(\x)=\delta^{(3)}(\x)$, and so the dual measure is simply the usual position space measure $d^3x$. However this is not the case for other theories. As we shall see, not only do these definitions lead to a self-consistent closure of the quantization procedure, but when we examine their implications for theories representing UV dimensional reduction, we will find that the position space measure has the same Hausdorff dimension in the UV as momentum space. This
result is far from trivial.

It is also interesting that this dual measure can be used to set up an inner product fully consistent with Section~\ref{postulate}.
(We again note that we keep the original coordinates in ${\bf x}$ and ${\bf p}$ in our transform, so that some of the subtleties pointed out in ~\cite{calcagni1,calcagni2} are not applicable to our construction.) Let
\be
(\phi,\psi)=\int d\overline\mu_x(\x) \phi^\star i \overleftrightarrow\partial_0 \psi\ .
\ee
Then, an orthonormal basis satisfying
\be
(e_\p,e_{\p'})=\delta_{\overline\mu}(\p-\p')
\ee
is supplied by:
\be
e_\p = e^{-ip\cdot x}
\ee
agreeing  with the basis used in expansion (\ref{expansion}). In addition, bearing in mind (\ref{norm}), we have:
\be
\langle 0 |\phi(x)|\p\rangle = e^{-ip\cdot x}
\ee
showing the self-consistency of the whole framework.

\subsection{First remarks on the meaning of dimensionality}\label{remarks}
Even though the position space structures proposed in Section~\ref{inner} are not needed for our main calculation,
illuminating insights can be gleaned from them when later in this paper we examine concrete models.
We should, however, beware of the multiple senses in which the concept of dimensionality is used to avoid confusion.
As explained in~\cite{DSRmeasure} we can eschew the concept of spectral dimension in favour of the Hausdorff dimension
of the measure in the linearizing frame. Dimensionality is then associated with the  power of the radial coordinate appearing in the measure (we will review examples of this in Sections~\ref{applications1} and ~\ref{applications2}). This is true both in position and momentum space. In momentum space this definition reflects how the number of modes
scales as we consider larger and larger momentum magnitudes. More generally it reflects  the scaling of the measure
under a dilation~\cite{calcagni1}.

We stress that in our constructions we keep the original coordinates $\x$ and $\p$ (here assumed to be 3 dimensional, but we shall
relax this assumption later in the paper). Therefore
the actual number of variables in cartesian  coordinates (or the number of angles in polar coordinates) is irrelevant
for the dimensionality, and stays constant.  Likewise the unit-dimensionality of the measure remains constant and is always
$L^4$ in position space or $L^{-4}$ in momentum space. The fact that the Hausdorff dimension changes can be reconciled with this fact
by recalling that there is invariably a length scale in the MDR, and a power of this dimensionful parameter makes up the difference.

Finally we make an important point regarding the Fourier transform and the delta function, which will be essential for our
calculations in Section~\ref{applications1} and~\ref{applications2}. Since we identify dimensionality via the power of the radial
coordinate when the
measure is written in polar coordinates, we should define a $D$ dimensional delta function $\delta ^D(\mathbf w)=
\delta^D(w)$ (with $w=|\mathbf w|$) via:
\be
1=\int d\mu(\mathbf w) \delta^D (\mathbf w)\propto \int dw\, w^{D-1}\delta^D(w)
\ee
where $\mathbf w$ may be either position or momentum variables, and where we assume that $D$ is the dimensionality of the
measure $d\mu$.  Therefore the $D$ dimensional delta function (where $D$ need not be an integer) is given by:
\be\label{deltaD}
\delta ^D(\mathbf w)= \delta^D(w)\propto \frac{\delta(w)}{w^{D-1}}
\ee
regardless of the number of components in $\mathbf w$ (which we shall assume fixed, even if the Hausdorff dimension runs). 
This delta function can be Fourier transformed, but just
as we keep the original variables, the Fourier transform involves $e^{i\p\cdot\x}$ where $\p$ and $\x$
are the original coordinates (here assumed to be 3 dimensional). However, it is still true that:
\be\label{deltaCENTRAL}
\delta^D({\bf x})=\int d\mu(\p) e^{i{\mathbf p}\cdot {\mathbf x}},
\ee
a result we shall prove here. 
Dropping numerical factors throughout, the right hand side of (\ref{deltaCENTRAL}) is proportional to:
\bea\label{inter}
\int dp p^{D-1}\int d(\cos\theta)e^{ipr\cos\theta}
&\propto&\int dp\, p^{D-2}\frac{e^{ip r}}{r}\nonumber\\
&\propto&
\frac{1}{r}
\frac{ d^{D-2}\delta(r)}{dr^{D-2}},
\eea
where $r=|\mathbf x|$ and $p=|\mathbf p|$.
Using the identities generated by
\be
\delta ^D(r)\propto
\frac{\delta (r)}{r^{D-1}}\propto
\frac{ d^{D-1}\delta(r)}{dr^{D-1}},
\ee
we can see that (\ref{inter}) is proportional to
$\delta ^D(r)=\delta^D(\x)$, proving (\ref{deltaCENTRAL}).

\section{Vacuum fluctuations under a non-trivial measure}\label{vac-flucts}
We now evaluate the power spectrum of vacuum quantum fluctuations for all theories that can be quantized according to
this procedure. It will be useful to define a ``covariant'' power spectrum (i.e. a power spectrum associated with the
deformed  measure), since our results will then take a
universal form, whereas the conventional power spectrum will be model-dependent.
We also define the dimensionless power spectrum and relate it to the covariant power spectrum. This will lead to
a universal relation between the spectral index and the Hausdorff dimension of momentum space.

\subsection{Vacuum fluctuations}
In order to study the power spectrum of vacuum fluctuations we consider the two-point function $\langle 0 |\phi (\x,0)\phi(\y,0)|0\rangle$. The power spectrum is defined via the expansion in Fourier modes
\be
\langle 0 |\phi (\x,0)\phi(\y,0)|0\rangle = \int \frac{d\p}{(2\pi)^3} \, P_\phi (\p) e^{i\p \cdot (\x-\y)}
\ee
which, for theories with statistical homogeneity and isotropy, can be written as:
\be\label{fulldefP}
\langle 0 |\phi (\x,0)\phi(\y,0)|0\rangle=\int \frac{dp}{2\pi^2} \, p^2 P_\phi (p) \frac{\sin pr}{pr}
\ee
where $r=|\x-\y|$. More simply we may identify the power spectrum from:
\be\label{convP}
\langle 0 |\phi^2 (x)|0\rangle= \langle 0 |\phi^2 (0)|0\rangle := \int \frac{d\p}{(2\pi)^3} \, P_\phi (\p)\ .
\ee
We find it pertinent to define an alternative ``covariant'' power spectrum, $ \overline P_\phi$, i.e. a power spectrum with respect to the covariant  on-shell momentum measure. This may be obtained from:
\be\label{barP}
\langle 0 |\phi ^2(x)|0\rangle:=\int d\overline\mu(\p)\, \overline P_\phi (\p)\; .
\ee
to be compared and contrasted with (\ref{convP}).

The covariant power spectrum  of the vacuum fluctuations in all theories that can be quantized as in Section~\ref{formalism} has a very simple universal form. By inserting (\ref{expansion}) into (\ref{barP}) and using (\ref{post}) and $\langle 0|0\rangle=1$, it is straightforward to prove the remarkable general result:
\be\label{!!}
\overline P_\phi (\p)=1\,,
\ee
which shows that the covariant power spectrum is {\it independent} of the actual measure in momentum space. The simplicity
and generality of this result would not be obvious in terms of the standard power spectrum.
The relation between the covariant and the standard power spectrum (obtained by comparing (\ref{barP}) and (\ref{convP})) is model dependent, and so the result for $P_\phi(\p)$ is also model-dependent, as we shall see in several cases later in this paper.
For isotropic theories, comparison between  (\ref{isomu}) and (\ref{convisomu}) leads to:
\be
\overline P_\phi(p)=
\frac{p^2} {(2\pi)^3\overline\mu(p)} P_\phi(p)
\ee
and so:
\be
P_\phi(p)=\frac{(2\pi)^3\overline\mu(p)}{p^2}.
\ee
For the ordinary QFT with undeformed measure (cf. (\ref{convisomu})) we find the relation
\be
\overline P_\phi=2E P_\phi
\ee
and so recover the well-known result:
\be
P_\phi(p) =\frac{1}{2E_p} \approx
\frac{1}{2p}
\ee
(where the near-massless condition was used).

\subsection{The dimensionless power spectrum and scale-invariance}\label{dimlessPsec}
We now translate our results to the notation used in cosmological perturbation theory.
We will also allow for the original number of spatial dimension to be
an arbitrary number, $D$, so as to make some general points here and later in the paper.
The quantity more often used in cosmology is the dimensionless power spectrum of the curvature fluctuation $\zeta$.
This is usually second quantized for modes inside the horizon; the vacuum fluctuations are then followed as the modes
leave the horizon. Although the dynamics and quantization of $\zeta$ and $\phi$ are the same, their unit-dimensions are different, since:
\be\label{zetaphi}
\zeta\propto {\sqrt G}\phi\propto L_P^{\frac{D-1}{2}}\phi.
\ee
In position space $\zeta$ is dimensionless in any number of dimensions, and so are its correlators.
The dimensionless power spectrum is obtained by multiplying the power spectrum of $\zeta$ by the
by appropriate powers of $p$ so as to render it dimensionless. By definition we thus have:
\be\label{fulldefdimlessP}
\langle 0 |\zeta (\x,0)\zeta(\y,0)|0\rangle: =\int \frac{dp}{p} \, {\cal P}_\zeta (p) \frac{\sin pr}{pr}
\ee
or more simply:
\be\label{dimlessP}
\langle 0 |\zeta^2 (x)|0\rangle= \langle 0 |\zeta^2 (0)|0\rangle := \int \frac{dp}{p} \,{\cal  P}_\zeta (p)\ .
\ee
In general the amplitude $A$ and spectral index $n_S$ can be read off from
\be
{\cal  P}_\zeta (p)=A^2\left(\frac{p}{p_0}\right)^{n_S-1}\; ,
\ee
where $p_0$ is a preferred wavenumber (required for matching the unit-dimensions).
Scale-invariant fluctuations do not have a $p_0$ and so require $n_S=1$.

Using (\ref{dimlessP}), (\ref{zetaphi}) and  (\ref{barP}) (and assuming isotropy, i.e. (\ref{isomu}))
leads to:
\be
{\cal P}_\zeta (p) \propto   G p{\overline \mu}(p){\overline P}_\phi(p)
\ee
(where as usual we have dropped numerical factors, which may depend on $D$).
Therefore  (\ref{!!}) implies the general result:
\be\label{eq:generaldimensionlessP}
{\cal P}_\zeta (p) \propto  G p{\overline \mu}(p)\propto L_P^{D-1}p{\overline \mu}(p).
\ee
Note that whatever the dimensionality of momentum space the measure $\bar\mu$ will have the appropriate
dimensions to make this expression
dimensionless. This is enforced by a power of the dimensionful scale $\ell$ appearing in the MDRs,  defining the UV
scale (cf. comments made in Section~\ref{remarks}). We stress that $\ell$  need not be $L_P$.

We can now prove a general result relating $n_S$ and the Hausdorff dimension of energy-momentum space, $d_H$.
As we shall see case by case we can always write the measure in the UV limit in the form:
\be
d\mu\propto \ell ^{d_H-1-D} dE\, E^{D_E-1}dp\, p^{D_p-1}
\ee
with $d_H=D_E+D_p$, and with the factor in $\ell$ ensuring the the original unit-dimensions of $d\mu$ are kept.
Using (\ref{onshellmu}) and assuming $E\approx p$  we thus have:
\be
d\overline\mu\propto dp\, \ell ^{d_H-1-D} p^{d_H-3}
\ee
so that  (\ref{eq:generaldimensionlessP}) implies:
\be
{\cal P}_\zeta (p) \propto {\left(\frac{L_P}{\ell}\right)}^{D-1}  (\ell p)^{d_H-2}.
\ee
Therefore, for {\it all} theories that can be fitted into our framework we have the relation:
\be\label{nsgeneral}
n_S=d_H-1.
\ee
The general condition for scale-invariance is
\be\label{scinvgeneral}
d_H=2.
\ee
(i.e. a two-dimensional energy-momentum space) and when this happens the
amplitude is:
\be
A\sim{\left(\frac{L_P}{\ell}\right)}^{\frac{D-1}{2}}
\ee
so that if $D=3$ we should require $\ell\sim 10^5 L_P$.
For the rest of this paper we will examine how these general conditions come about
for different reasons at the intermediate steps, on a case by case basis,
both for theories with and without a preferred frame.

We stress that running  dimensionality with a two-dimensional UV limit is very different from just postulating ordinary
dispersion relations and a 2D universe. Using the results above it is straightforward to show that the latter would lead
to ${\cal P}_\zeta\sim 1$, i.e. a scale-invariant spectrum with an amplitude of order 1. Having a hierarchy between $\ell$ and $L_P$ is essential for getting the amplitude right.

\section{Applications I: LIV theories}\label{applications1}
We now examine the implications for the various alternative theories that may be fitted into this framework, starting with those with explicit Lorentz invariance violation (LIV) and a preferred frame. We first review the sub-class of LIV theories for which cosmological fluctuations have already been studied \cite{DSRflucts1,muko,DSRflucts2,DSRrainbow,DSRmeasure}, showing how our results correspond to established facts (specifically as reworked in \cite{DSRmeasure}). We then consider the general LIV case.

\subsection{Horava-Lifshitz type theories}\label{HL}
Let us consider a LIV theory based on MDRs $\Omega(E,p)=0$, with:
\begin{equation}\label{livMDR}
\Omega= E^{2}+p^{2}+\ell_{t}^{2\gamma_{t}}\hat E^{2(1+\gamma_{t})}-\ell_{x}^{2\gamma_{x}}\hat p^{2(1+\gamma_{x})} - m^2 \, .
\end{equation}
whose UV limit is:
\begin{equation}\label{livMDR}
\Omega\approx \ell_{t}^{2\gamma_{t}}\hat E^{2(1+\gamma_{t})}-\ell_{x}^{2\gamma_{x}}\hat p^{2(1+\gamma_{x})} - m^2 \, .
\end{equation}
and undeformed measure. This is the expression of the theory in the MDR frame (denoted by a hat), but as stated,
we shall re-express the theory in linearizing units, with standard dispersion relations and a  deformed measure.
When  $\gamma_t=0$ we can perform the fluctuations calculation equally well in the linearizing~\cite{DSRmeasure,DSRrainbow}
or the MDR frame~\cite{DSRflucts1,DSRflucts2},
with little difference in labour.  We recap the main results
in~\cite{DSRmeasure}, and derive them within our framework, with a few interesting additions.

When $\gamma_t=0$, in the UV limit the linearizing units  are given by:
\bea
E&=&\hat E\\
p&=&\ell_{x}^{\gamma_{x}}\hat p^{(1+\gamma_{x})}.
\eea
In terms of these, the dispersion relations are quadratic in the UV but the
the measure becomes:
\be\label{dmuHL}
d\mu\propto dE\, p^{D_p-1}dp
\ee
with
\be
D_p=\frac{D}{1+\gamma_x}.
\ee
The Hausdorff dimension of energy-momentum space in the UV limit is therefore:
\be
d_H= 1+\frac{D}{1+\gamma_x}
\ee
in agreement with the well-known result found using the concept of spectral dimension.

Given (\ref{dmuHL}) and (\ref{onshellmu}), the covariant on-shell measure is:
\be\label{dbarmuHL}
d\overline \mu \propto \frac{p^{D_p-1}dp}{2E_\p}
\ee
from which it follows that:
\be
\delta_{\overline \mu} (\p) \propto 2E_p\, \delta^{D_p}(\p)
\ee
where  $\delta ^{D_p}(\p)$ is defined by (\ref{deltaD}).
Our basic postulate, Eq.~(\ref{post}), is then just the standard quantization relation
\be\label{comsHL}
[a(\p),a^\dagger (\p')]\propto  2 E_\p\, \delta^{D_p}(\p-\p')
\ee
in $D_p$ spatial dimensions (a statement which is non-trivial if $D_p$ is non-integer).
This leads to the well-known result for the power spectrum of these theories~\cite{DSRmeasure,DSRrainbow}. Since:
\be
\overline\mu(p)\propto \frac{p^{D_p-1}}{E_p}
\ee
the relation between the covariant and the standard power spectrum is:
\be
\overline P_\phi\propto  E_p \, p ^{3-D_p} P_\phi ,
\ee
and so:
\be
P_\phi(\p) \propto \frac{p^{D_p-3}}{ E_p}.
\ee
The dimensionless power spectrum is  therefore:
\be
{\cal P}_\phi(p)\propto \frac{p^{D_p}}{E_p}\approx p^{D_p-1}
\ee
(where we have assumed $E_p\gg m$) and so $n_S=D_p$, or:
\be
n_S=d_H-1
\ee
and
HL theory with $\gamma_x=2$, $D_p=1$, and $d_H=2$ is the only case for which there is
scale-invariance.

Based on Section~\ref{inner} we can add a new result to HL theory. We can show that the picture of dimensional reduction found in momentum space extends to position space. As explained in Section~\ref{inner}, the quantization condition (\ref{comsHL}) is equivalent to the standard equal-time commutation relations in position space (in the linearizing frame) if we use the modified spatial delta function:
\be
\delta_{\overline\mu_x}(\x)\propto \int d(\cos\theta)\, dp\, p^{D_p-1} e^{i\p\cdot \x}
\ee
(cf. (\ref{deltax}) and (\ref{dbarmuHL})). But using  the non-trivial result (\ref{deltaCENTRAL}) (see proof following
Eq.~(\ref{deltaCENTRAL})), we have that
\be
\delta_{\overline\mu_x}(\x)\propto \delta^{D_p}(\x)
\ee
so that the spatial delta function
and position space spatial measure are also $D_p$ dimensional. Therefore we find the important result that for the HL model space and time mimic momentum and energy in terms
of dimensionality.

\subsection{LIV case when $\gamma_t\neq 0$}\label{gtneq0}
If $\gamma_t\neq 0$ we should start our calculation by adopting  linearizing units.
Transitioning to field theory in the  MDR frame would involve higher than second order derivatives in time, leading to well
known problems. Linearizing the theory first, and only then translating MDRs into field theory circumvents these problems.

The linearizing units are:
\bea
E&=&\ell_{t}^{\gamma_{t}}\hat E^{(1+\gamma_{t})}\\
p&=&\ell_{x}^{\gamma_{x}}\hat p^{(1+\gamma_{x})}
\eea
so that if in the original frame the MDRs are (\ref{livMDR}) and the integration measure is trivial, in
the linearizing frame the dispersion relations are  quadratic but the measure is:
\be
d\mu\propto E^{D_E-1}dE\, p^{D_p-1}dp \label{eq:measureLIVgeneral}
\ee
with:
\bea
D_E&=&\frac{1}{1+\gamma_t}\\
D_p&=&\frac{D}{1+\gamma_x}.
\eea
The Hausdorff dimension of energy-momentum space is therefore $d_H=D_E+D_p$, in agreement with
the result obtained using the spectral dimension (see~\cite{Sotiriou:2011aa}).

The on-shell measure is now:
\be
d\overline \mu\propto E^{D_E-2} p^{D_p-1}dp
\ee
from which it follows that:
\be
\delta_{\overline \mu} (\p) \propto E^{2-D_E}_\p\, \delta^{D_p}(\p)
\ee
where $\delta^{D_p}(\p)$ is the delta function in $D_p$ dimensions as defined in (\ref{deltaD}).
Eq.~(\ref{post}) then amounts to postulating:
\be\label{post0}
[a(\p),a^\dagger (\p')]\propto E^{2-D_E}_\p\, \delta^{D_p}(\p- \p') .
\ee
As explained in Section~\ref{inner}, this is equivalent to the standard equal-time commutation relations in position space
if we use the modified spatial delta function:
\be\label{general-delta}
\delta_{\overline\mu_x}(\x)\propto \int dp\, p^{D_p-1} E_\p^{D_E-1} e^{i\p\cdot \x}.
\ee
Since we are working in the UV, and $E\approx p$ in this regime, this is equivalent to
\be
\delta_{\overline\mu_x}(\x)\propto \delta^{D_p+D_E-1}(\x).
\ee
In other words space (as opposed to space-time) has acquired dimensionality
\be
D_{\vec x}= D_p+D_E-1.
\ee
The full space-time dimensionality is therefore
\be
D_x=1+D_{\vec x}= D_p+D_E
\ee
i.e.: the same as that of energy-momentum space. However the split between energy and momentum and that between space and
time no longer mimic each other.

This is a new result arising from our framework, but more importantly we can now evaluate the power spectrum of vacuum
quantum fluctuations for the wider class of LIV theories.
Given that:
\be
\overline\mu(p)\propto E_\p^{D_E-2}p^{D_p-1}
\ee
the relation between the covariant and the standard power spectrum is now
\be
\overline P_\phi\propto  E^{2-D_E}_\p p ^{3-D_p} P_\phi ,
\ee
so that:
\be
P_\phi(\p) \propto p^{D_p-3} E^{D_E-2}_\p.
\ee
The dimensionless power spectrum is
\be
{\cal P}_\phi(p)\propto E_p^{D_E-2}p^{D_p}\approx p^{D_E+D_p-2}
\ee
where the last approximation assumes $E\gg m$. Therefore
\be
n_S=D_E+D_p-1=d_H-1
\ee
and scale-invariance is equivalent to:
\be
d_H=D_E+D_p=2, \label{eq:generalLIVscaleinvariance}
\ee
as before, generalizing the result previously obtained for HL theory for the whole class of LIV theories.
This is true for modes inside the horizon (or ignoring gravity). The role of gravity will be addressed in
Section~\ref{conclusions}.

\section{Applications II: Theories without a preferred frame}\label{applications2}
We now consider how our general results arise through somewhat different intermediate steps in the case of theories
without a preferred frame. We consider the case of non-local special-relativistic theories, and of a DSR-relativistic theory with de Sitter momentum space, for which Lorentz transformations are deformed. 

\subsection{Non-local Lorentz invariant theories}
Following some previous literature \cite{Nesterov:2010yi} we will call these models ``non-local Lorentz invariant" theories.
They are the simplest theories with MDRs which do not introduce preferred frames and will allow us to
illustrate a technique that will be used in the more complex case of de Sitter-DSR.

In these models the MDR  is given by:
\be
E^{2}-p^{2}+\ell^{2\gamma}(\hat E^2-\hat p^2)^{\gamma+1}=m^2,
\ee
which in the UV limit reads:
\be
\ell^{2\gamma}(\hat E^2-\hat p^2)^{\gamma+1}=m^2.
\ee
The measure is just the standard one:
\be
d\mu=d\hat E \,d^{D}\hat p = d\hat E\, d\hat p\, \hat p^{D-1}.
\ee
In order to find the UV linearizing units $\{E,p\}$ we introduce hyperbolic coordinates in momentum space, defined by:
\bea
\hat E&=&\hat r \cosh \hat \varphi\nonumber\\
\hat p&=&\hat r \sinh \hat \varphi\, .
\eea
In these coordinates the UV dispersion relation reads:
\be
\ell^{2\gamma} \hat r^{2(\gamma+1)}=m^2,
\ee
whilst the measure becomes:
\be
d\mu\propto \hat r^D (\sinh\hat\varphi)^{D-1} \,d\hat r\,d\hat\varphi\, .
\ee
We can define linearizing hyperbolic coordinates $\{r,\varphi\}$, such that the dispersion relation becomes trivial:
\be
 r^{2}=m^2,
\ee
simply by setting:
\be
r\equiv \hat r^{1+\gamma}
\ee
with
\bea
 E&=& r \cosh  \varphi\nonumber\\
 p&=& r \sinh  \varphi\, .
\eea
At this point $\varphi$ could be anything, but under closer scrutiny we notice that by keeping the original $\hat \varphi$ we introduce a pathological coordinate system. This is signalled by the divergence or collapse to zero of the integration measure when $E=p$, as evident by reverting back to cartesian coordinates. Another pathology is noted by imposing the mass shell condition in these units: whereas this constraint should reduce the dimensionality by one, it appears to change it from $D+1/(1+\gamma)$ to $D$.

These pathologies can be avoided by requiring that $\varphi$ should be such that by reverting to cartesian coordinates
$\{E,p\}$ the measure should read:
\be
d\mu\propto dE \, E^{D_E-1} \, dp \, p^
{D_p-1}
\ee
for some $D_E$ and $D_p$, rather than acquire factors of powers of $-E^2+p^2$. This requires:
\be
(\cosh\varphi)^{D_E-1} \,(\sinh\varphi)^{D_p-1}=(\sinh\hat\varphi)^{D-1}\frac{d\hat\varphi}{d\varphi}
\ee
with
\be
D_E+D_p-2=\frac{D-1-2\gamma}{1+\gamma} \label{eq:DSRMinkowskidimensions}
\ee
equivalent to
\be\label{dhminkDSR}
d_H=D_E+D_p=\frac{1+D}{1+\gamma}.
\ee
Such a construction does not uniquely specify the new variable $\varphi$ (or the separate values of $D_E$ and $D_p$), but
$d_H=D_E+D_p$ is fixed given $D$ and $\gamma$, so that the dimensionality of the measure $d\mu(E,p)$ is fully determined.  Once we go on-shell this does not matter, but we could choose $D_E=1$ for reasons of symmetry.
Then:
\be
(\sinh\varphi)^{d_H-2}=(\sinh\hat\varphi)^{D-1}\frac{d\hat\varphi}{d\varphi}
\ee
which in the UV (approximating the $\sinh$ by the dominant exponentials) leads to
\be
\varphi =\frac{(1+\gamma)D}{D-\gamma}\hat\varphi\, ,
\ee
in terms of which the measure becomes:
\be
d\mu=dE\, dp\, p^{d_H-2}.
\ee

One can now follow the same steps after eq. (\ref{eq:measureLIVgeneral}) for LIV theories. In particular the on-shell measure is now:
\be
d\overline\mu=dp\, p^{d_H-3},
\ee
in the massless approximation.
We arrive to the same general result for $n_S$ as (\ref{nsgeneral}) and condition for scale-invariance as (\ref{scinvgeneral}).
Specifically we require:
\be
d_H=\frac{1+D}{1+\gamma}=2
\ee
which implies  $\gamma=1$ in $3+1$ dimensions.

\subsection{de Sitter-DSR}\label{sec:DSR}

The most studied theory with MDRs which does not introduce a preferred frame is de Sitter-DSR.
We consider a rendition of the de Sitter momentum-space model in $D+1$ dimensions, which in ``bicrossproduct coordinates'' \cite{Majid:1994cy,Arzano:2014jua} is characterized by the MDR:
\be
\Omega=\mathcal C_\ell-m^2\equiv\frac{4}{\ell^2}\sinh^2\left(\frac{\ell \hat E}{2}\right)-e^{\ell \hat E}\hat p^2-m^2
\ee
and momentum-space measure:
\be
d\mu = e^{D \ell \hat E} \hat p^{D-1} d\hat E d\hat p .
\ee
The linearizing units are
\bea
E&=&\frac{2\sinh\left(\ell\hat E/2\right)}{\ell}\nonumber\\
p&=&\hat p \,e^{\ell \hat E/2},\label{eq:LinearizingUnitsdS}
\eea
which in the UV (defined as $\hat E\rightarrow \infty$, $\hat p\rightarrow 1/\ell$) become:
\bea
E&\approx&\frac{e^{\ell \hat E/2}}{\ell}\nonumber\\
p&=&\hat p \,e^{\ell \hat E/2}.\label{eq:LinearizingUnitsdSUV}
\eea
In terms of these the momentum-space  measure reads:
\be
d\mu\propto  E^{D-1} p^{D-1} dE dp.\label{eq:measuredS}
\ee
From this measure we can infer the energy-momentum space dimensionality $d_{H}=2D$.

We can now carry out the quantization programme spelled out in Section~\ref{formalism}.
The on-shell measure for our de Sitter DSR is:
\be
d\bar\mu\propto E_{\p}^{D-2} p^{D-1} dp,
\ee
so that the associated delta function in Fourier space is given by:
\be
\delta_{\bar\mu}(\p)\propto E_{\p}^{2-D}\delta^D(\p).
\ee
Given the form of this delta function we know that
the annihilation and creation operators commutation relations are:
\be\label{DSRcom}
[a(\p),a^\dagger (\p')]\propto E_{\p}^{2-D} \delta^{D}(\p- \p') .
\ee
where $\delta^D$ is the standard delta function in $D$ dimensions (defined according to the procedure described at the end 
of Section~\ref{remarks}). As discussed before, in the case of DSR this relation should only be applied on the vacuum, if we do not want to surreptitiously introduce a preferred frame. In DSR momenta must exhibit a {\it non-abelian} composition rule, which would be reflected in a deformed commutator of the type $a(\p)a^\dagger (\p')-a^\dagger (\p') a(\tilde{\p}(\p,\p')) \propto E_{\p}^{2-D} \delta^{D}(\p- \p')$, where $\tilde{\p}(\p,\p')$ is a non-trivial function of $\p$ and $\p'$ which reduces to $\p$ in the limit of vanishing $\ell$ \cite{arzano-marciano,arzanoQFT,Arzano:2013sta}. The commutator (\ref{DSRcom}) knows about the whole Fock space of the theory and thus adopting an undeformed one we are making a strong statement about the statistics of the model, which introduces a preferred frame in the multi-particle sector.
However, in~\cite{giulia} we explicitly prove that the quantization procedure proposed in~\cite{arzano-marciano,arzanoQFT,Arzano:2013sta}, constructed to be compatible with the deformed composition of momenta induced by the symmetries of the de Sitter momentum space, is nothing but that proposed here at the level of the second order action.

Given this quantization procedure the dimensionless power spectrum can be found using \eqref{eq:generaldimensionlessP}, with $\bar\mu(p)\propto E_{\p}^{D-2} p^{D-1}$:
\be
\mathcal P_\phi(p)\propto E_{\p}^{D-2} p^{D}\approx p^{2D-2}.
\ee
Therefore scale invariance is not achieved; rather 
\be
n_S=2D-1,
\ee
so that for $D=3$ we have $n_S=5$, i.e. a very blue spectrum.

\subsection{de Sitter-DSR with $\gamma\neq 0$}
Let us now analyze a de Sitter momentum-space characterized by a UV dispersion relation of the form:
\be
\ell^{2\gamma} \mathcal C_\ell^{1+\gamma}=m^2.
\ee
This case was studied in~\cite{Amelino-Camelia:2013cfa} and is known to produce a UV dimension of 2 if $\gamma=2$.
We can start the linearization procedure using the coordinates $\{E,p\}$ defined in (\ref{eq:LinearizingUnitsdS}), so that the dispersion relation becomes
\be
\ell^{2\gamma} (E^2-p^2)^{1+\gamma}=m^2
\ee
with measure given by Eq. (\ref{eq:measuredS}). Once this is done the situation is completely analogous to that described for  ``non-local Lorentz invariant"  theories  (the only difference is that the measure has an extra $E^{D-1}$ factor). We start by rewriting 
de Sitter-DSR expressed in linearizing coordinates in the associated hyperbolic coordinates:
\bea
 E&=& r \cosh \varphi\nonumber\\
 p&=& r \sinh  \varphi\, ,
\eea
so that the MDR becomes
\be
\ell^{2\gamma} r^{2(\gamma+1)}=m^2,
\ee
with volume measure:
\be
d\mu=dr\, d\varphi\, r^{2D-1}(\cosh\varphi)^{D-1}
(\sinh\varphi)^{D-1} \, .
\ee
Linearizing for $\gamma\neq 0$ can be achieved by
\be
\tilde r\equiv r^{1+\gamma}
\ee
and 
\be
(\cosh\tilde\varphi)^{D_E-1} \,(\sinh\tilde \varphi)^{D_p-1}=(\cosh\varphi)^{D-1}(\sinh\varphi)^{D-1}\frac{d\varphi}{d\tilde \varphi}
\ee
with
\be
D_E+D_p-2=\frac{2(D-1-\gamma)}{1+\gamma},
\ee
which is equivalent to:
\be\label{dhDSR}
d_H=D_E+D_p=\frac{2D}{1+\gamma},
\ee
(this will turn out to be the Hausdorff dimension of the space). 
Among the possible redefinitions of $\varphi$ which avoid  pathologies  we can choose the one which mimics the original space in giving equal dimensions to energy and momentum space. This amounts to defining:
\bea
d\varphi\, (\cosh\varphi)^{D-1}  (\sinh\varphi)^{D-1} =&&\nonumber\\
d\tilde \varphi\, (\cosh\tilde\varphi)^{\frac{D}{1+\gamma}-1}  (\sinh\tilde\varphi)^{\frac{D}{1+\gamma}-1} d\tilde\varphi&&
\eea
which in the UV (approximating the sinh by the dominant exponentials) leads to
\be
\tilde \varphi =\frac{d_H-1}{2D-\gamma}\varphi\, .
\ee
In terms of  variables
\bea
\tilde E&=& \tilde r \cosh\tilde  \varphi\nonumber\\
 \tilde p&=& \tilde r \sinh  \tilde \varphi\,
\eea
the measure is:
\be
d\mu=d\tilde E\, d\tilde p\, \tilde  p^{\frac{D}{1+\gamma}-1} \tilde E ^{\frac{D}{1+\gamma}-1},
\ee
so that indeed $d_H$ given by (\ref{dhDSR}) is the Hausdorff dimension of the space. 
As with non-local Lorentz theories a more general non-pathological $\tilde\varphi$ could be defined, with the
same $d_H$ but different $D_E$ and $D_p$.

The quantization of this theory can be performed just as with DSR, with the same pitfalls and provisos described in Section~\ref{sec:DSR}. 
One can follow the same steps following eq. (\ref{eq:measureLIVgeneral}) for LIV theories to quantize the theory and
work out the spectrum of vacuum fluctuations.  The on-shell measure in the UV limit is:
\be
d\overline\mu=d\tilde p\, \tilde p^{d_H-3},
\ee
so that following the same steps we find:
\be
n_S=d_H-1=\frac{2D}{1+\gamma}-1.
\ee
Therefore once again scale-invariance is achieved when $d_H=2$, which now amounts to choosing $\gamma=2$ when $D=3$.

\section{Discussion on dimensionality}
In view of our findings, in this Section we collect some general results on the dimensionality of momentum space and on the possible
position space dual, in theories with and without frame dependence.

\subsection{A quantum vacuum dimension?}
The fact that quantum-gravity research appears to lead inevitably to spacetime quantization renders unavoidable the task of characterizing the dimensionality of such ``quantum spacetimes''. This highly non-trivial task has so far been handled by resorting to the notion of spectral dimension. We feel that this approach has several shortcomings, at least regarding its applicability to the most sophisticated pictures of quantum spacetime which have appeared in the literature. Our main concern is that the spectral dimension is not defined directly in terms of observable quantities. Rather, it relies on the evidently unphysical step of Euclideanization of the theory, with the further dubious addition of a fictitious diffusion process in that space.

Naturally, for some theories the spectral dimension will be numerically correct, even though its definition does not have direct operative physical meaning. For example, in  HL theories it has been noted \cite{Husain:2013zda, Santos:2015sva} that the dimensionality of spacetime obtained via the spectral dimension algorithm does match the indications on dimensionality that emerge by studying the  Stefan-Boltzmann law, $\rho\propto T^\alpha$, in those theories.
However, outside of the HL context there is at present no evidence that the spectral dimension truly acquires a physical meaning.

On the basis of our work we argue that the spectral index of vacuum fluctuations could provide a good
candidate for an operationally measurable concept of dimensionality of spacetime.
Indeed we have found that for all theories that can be
fitted into our scheme the spectral index and the momentum space Hausdorff dimension are related by:
\be
n_S=d_H-1.
\ee
Moreover, as already noted for LIV theories (and discussed in more generality in the next subsection), the dual position space defined in  Section ~\ref{inner} and momentum space share the same dimensionality.
This does not depend on the details of the theory, for example any HL type theory with $\gamma_t$ and $\gamma_x$ leading to
the same $d_H$ produces the same $n_S$. It also applies to DSR theories
regardless of their subtleties in the multiparticle sector, since we only need to evaluate
the commutator (\ref{post}) as applied on the vacuum, as explained before.

Vacuum fluctuations are always
observable {\it in principle}. They could also already have been seen in the sky as a relic from the dimensionally reduced phase.
The latter statement is stronger, and 
relies on a complex ``transfer function'', dependent on the theory of gravity and how the modes end up
outside the horizon at the end of this primordial phase. The dependence on these details is probably weaker if the primordial phase
is two-dimensional (as discussed in the conclusions). Regardless of this practical matter, the 
point remains that in principle such a ``quantum vacuum dimension'' could provide
an experimentally operational definition of the dimensionality of spacetime.

\subsection{The dimension of the position space dual measure}
In Section~\ref{inner} we proposed a dual measure in position space based on the principles of canonical quantization.
We have already considered its implications for LIV theories in Section~\ref{applications1}  but  have deferred until now
a general appraisal of the matter, including theories without a preferred frame.
In fact the argument following (\ref{general-delta}) presented for LIV theories
still applies formally to all theories without a preferred frame. From (\ref{deltax}) we still have
\be\label{deltax-general}
\delta_{\overline\mu_x}(\x)\propto \int dp\, p^{D_p-1} E_\p^{D_E-1} e^{i\p\cdot \x},
\ee
with $D_E+D_p=d_H$ as given by (\ref{dhminkDSR})
for non-local Lorentz invariant theories, with $D_E=D_p=D$ for DSR, and with $D_E+D_p=d_H$ as
given by (\ref{dhDSR}) for DSR with $\gamma\neq 0$. Just as in the LIV case, when $E_p\approx p$, the expression
(\ref{deltax-general}) becomes
\be
\delta_{\overline\mu_x}(\x)\propto \int d(\cos\theta)\, dp\, p^{d_H-2} e^{i\p\cdot \x},
\ee
and using the derivation resulting in (\ref{deltaCENTRAL}) we have for all the cases under consideration
(with LIV or not):
\be
\delta_{\overline\mu_x}(\x)\propto \delta^{d_H-1}(\x).
\ee
Therefore the dual measure has 1 time dimension and $d_H-1$ spatial dimensions, making up the same $d_H$
as for energy-momentum space. For example for de Sitter-DSR with $D=3$, energy-momentum
space is 6 dimensional in the UV limit, whereas
space is 5 dimensional and  space-time is 6 dimensional, the same as energy-momentum space.

The matching of the dimensions of spacetime and energy-momentum is even more remarkable considering that  only the total dimensionality of energy-momentum space, $d_{H}$, is always well defined (in general one can not assign separate dimensions to energy and spatial momentum; e.g. DSR with $\gamma\neq0$). In the spacetime picture we have that time and space can be separated, but only the space dimensionality is modified in the UV. Therefore there can never be a direct correspondence between the dimensions of energy and time, and between those of space and momentum. Yet, the numbers conjure to result in the same dimensionality for spacetime and energy-momentum space.

\section{Conclusions: General criteria for scale-invariance}\label{conclusions}
In this paper we set up a quantization framework for a large class of theories with deformed dispersion relations, with
and without preferred frames. We applied this procedure to the evaluation of the power spectrum of quantum vacuum fluctuations, and 
derived the following general results:
\begin{itemize}
\item The ``covariant'' power spectrum of the vacuum fluctuations is universally $\overline P(p)=1$. However, the conventional power spectrum $P(p)$ is model dependent.
\item The condition for the fluctuations to be scale-invariant is that the energy-momentum space should have Hausdorff dimension $d_H=2$
in the frame where the dispersion relations are linearized. 
\item More generally we have that the spectral index is related to the Hausdorff dimension of energy-momentum space as
$n_S=d_H-1$ for {\it all} theories.
\end{itemize}
Thus, we succeeded in relating the phenomenon of dimensional reduction to 2 in the UV, so ubiquitous in quantum gravity,
and a zeroth order benchmark in cosmological perturbation theory: scale-invariance. The amplitude of the fluctuations can also be tuned
to match the observed value by introducing a hierarchy between the Planck scale $L_P$ and the energy scale triggering the
UV dimensional reduction (as explained at the end of Section~\ref{dimlessPsec}).

These conclusions were drawn in flat space, but remain true for inside the horizon modes in an expanding universe.
In the absence of a theory of gravity associated with the large class of theories with MDRs considered here, we can 
simply set up the conditions to be satisfied by such a theory, should the scale-invariance be required to survive gravity
and the transition outside the horizon.  Specifically we can take cue from HL  theories as described in linearizing 
units~\cite{DSRmeasure}, for which the scale-invariant scenario is associated with the conformal coupling of gravity
to the fluctuations. In fact gravity drops out of the picture in the linearizing frame, which is why scale-invariance is universal 
(e.g. independent of the background equation of state) in the original frame, too. 

We propose that, for all theories with MDRs that can be fitted in our framework, the theory of gravity should be such that the second order action for fluctuations around a flat Friedmann model, as written in the linearizing frame, is:
\be
S_2=\int dk\, d\eta \left[\zeta'^2 +k^2 \zeta^2\right]
\ee
whenever the UV spectral dimension is $d_H=2$ (here $k$ is the comoving wavenumber, related to the physical one
by $k=p/a$, where $a$ is the expansion factor) .
This is just the usual action in the presence of gravity for radiation (which is conformally invariant), but here we posit that this 
conformal coupling is valid for all types of matter, as is the case for HL theory with $\gamma=2$. 
For HL theory, conformal coupling
in the linearizing frame is equivalent to Einstein gravity in the MDR frame (which was therefore labelled the Einstein
frame, in analogy with the Brans-Dicke terminology). However, this would not be the case for the other theories considered 
in this paper,  even
when $d_H=2$ and $n_S=1$. 
It would be interesting to investigate what the gravity theory might be in the MDR frame for all the other theories capable of generating scale-invariant fluctuations via MDRs. Specifically, one may wonder what the implications are for the zeroth order solution
with regards to the flatness and other problems usually solved by inflation, as well as the cosmological constant problem.
We defer this investigation to future work.

The fact that all the theories that can be fitted into our framework lead to the same observational prediction regarding the power spectrum 
should not worry us unduly. We have only established a universal connection between the zeroth
order target in observational cosmology---exact scale-invariance---and running of the dimensionality to two in the UV, assuming exact conformal invariance
of the gravitational coupling. This does not imply that these theories are observationally undistinguishable, once the devil in the detail
is examined.

Firstly, we do observe departures from exact scale-invariance, and the way these departures
are accomplished might be a strong discriminator between theories. In one specific case (that of theories modelled
on HL), we have previously tried to explain these departures applying the cavalier philosophy of the
cosmologist in the throes of model building~\cite{Amelino-Camelia:2013cfa,Barrow:2013gia,Barrow:2014fsa}.
But we can also speculate that these departures 
arise from a fundamental reason.
Specifically, the universe could start in a state where gravity existed in a limbo because it was only conformally coupled to all particles. 
Gravity and the Big Bang Universe are then relics from the mechanism responsible for breaking the fundamental conformal invariance. Such a mechanism should leave clear signatures in departures from scale-invariance in the primordial power spectrum.
We have voiced these speculations in~\cite{essay} and it is curious that they resonate with those 
in~\cite{tooftessay}.

Secondly, the bispectrum is likely to be a very powerful discriminator between theories. We have already pointed out that DSR
has a very non-trivial multi-particle sector, as opposed to theories with LIV. This could manifest itself in the bispectrum,
sensitive to the cubic action and the vertices of the theory. We speculate that the non-trivial addition rules of DSR will impart
a tell-tale signature upon the shape of the bispectrum. More generally, theories with and without LIV should leave a strong imprint into
the bispectrum and trispectrum of the primordial fluctuations. We are currently carrying out these calculations. 

Much work remains to be done, following on from this paper. We could for example explore scenarios where thermal, rather than
vacuum fluctuations, lead to the cosmological structures. For conventional theories, the spectrum of 
thermal fluctuations can be obtained from that of vacuum fluctuations simply by subtracting one power of $p$
(see, e.g.~\cite{Ferreira:2007cb}). Transposing this conclusion to theories with MDRs can be non-trivial, since thermal
states rely  on the multi-particle sector. But if the same conclusion holds, one would expect the rule 
$n_S=d_H-2$ for the spectral index of thermal fluctuations in the class of theories considered here.
Would  $d_H=3$ be the sweet spot theories should be aiming for?

\section{Acknowledgments}
We acknowledge support from the John Templeton Foundation. The work of MA was also supported by a Marie Curie Career Integration Grant within the 7th European Community Framework Programme. JM was further funded by an STFC consolidated grant and the Leverhulme Trust.

\end{document}